\documentclass[aps,prl,reprint,superscriptaddress,
amsfonts,amssymb,amsmath,a4paper,floats,floatfix]{revtex4-1}
\usepackage{graphicx}
\usepackage{siunitx}
\usepackage[dvipsnames]{xcolor}

\begin{document}


\newcommand{\TDEG}{two-dimensional electron gas}
\newcommand{\LL}{{\rm \scriptscriptstyle L}}
\newcommand{\RR}{{\rm \scriptscriptstyle R}}
\newcommand{\pdagger}{{\phantom \dagger}}


\title{Cavity-mediated coherent coupling between distant quantum dots}

\author{Giorgio\ Nicol\'i}
	\affiliation{Solid State Physics Laboratory, ETH Z\"urich, 8093 Z\"urich, Switzerland}

\author{Michael Sven\ Ferguson}
	\affiliation{Institute for Theoretical Physics, ETH Z\"urich, 8093 Z\"urich, Switzerland}

\author{Clemens\ R\"ossler}
	\affiliation{Infineon Technologies Austria, Siemensstra{\ss}e 2, 9500 Villach, Austria}
			
\author{Alexander\ Wolfertz}
	\affiliation{Solid State Physics Laboratory, ETH Z\"urich, 8093 Z\"urich, Switzerland}

\author{Gianni\ Blatter}
	\affiliation{Institute for Theoretical Physics, ETH Z\"urich, 8093 Z\"urich, Switzerland}
	
\author{Thomas\ Ihn}
	\affiliation{Solid State Physics Laboratory, ETH Z\"urich, 8093 Z\"urich, Switzerland}
	
\author{Klaus\ Ensslin}
	\affiliation{Solid State Physics Laboratory, ETH Z\"urich, 8093 Z\"urich, Switzerland}
	
\author{Christian\ Reichl}
	\affiliation{Solid State Physics Laboratory, ETH Z\"urich, 8093 Z\"urich, Switzerland}
	
\author{Werner\ Wegscheider}
	\affiliation{Solid State Physics Laboratory, ETH Z\"urich, 8093 Z\"urich, Switzerland}

\author{Oded\ Zilberberg}
	\affiliation{Institute for Theoretical Physics, ETH Z\"urich, 8093 Z\"urich, Switzerland}

\date{\today}


\begin{abstract}
Scalable architectures for quantum information technologies require to
selectively couple long-distance qubits while suppressing environmental noise
and cross-talk. In semiconductor materials, the coherent coupling of a single
spin on a quantum dot to a cavity hosting fermionic modes offers a new
solution to this technological challenge. Here, we demonstrate coherent
coupling between two spatially separated quantum dots using an electronic
cavity design that takes advantage of whispering-gallery modes in a
two-dimensional electron gas. The cavity-mediated long-distance coupling
effectively minimizes undesirable direct cross-talk between the dots and
defines a scalable architecture for all-electronic semiconductor-based
quantum information processing.
\end{abstract}

\maketitle


Quantum information technologies emerge as a promising solution to overcome
both the technological and the computational boundaries that limit standard
computers~\cite{nielsen_quantum_2010}. Quantum processing units operate with
qubits---quantum bits of information. They are realized using two-levels
systems and take advantage of the quantum principles of superposition and
entanglement of
states~\cite{nielsen_quantum_2010,loss_quantum_1998,clarke_superconducting_2008}.
These quantum properties can lead to a significant increase in our ability to
solve certain types of problems, with notable examples in the fields of
cryptography~\cite{shor_polynomial-time_1999} and simulation of quantum
many-body systems~\cite{abrams_simulation_1997}. Building a quantum computer
poses a multitude of challenges as many components need to work together in a
robust and scalable fashion. Numerous technologies are currently competing to
become the leading platform for quantum information processing~\cite{devoret2013,Awschalom2013,Rotta2017}.

Among them, semiconductor materials offer the possibility to encode qubits using artificial atoms embedded in a
two-dimensional electron gas---so called quantum dots~\cite{hayashi_coherent_2003,gorman_charge-qubit_2005,petta_coherent_2005}.
While single qubit operations via local control have been successfully
implemented, two-qubit entanglement requires a tunable coupling that is
difficult to achieve. Such coupling should be scalable, noise resistant, and
selective, requirements that become increasingly demanding as the density of
qubits is increased~\cite{Loss1998,devoret2013,Awschalom2013,Rotta2017}.  Dedicated coherent systems that mediate
tunable couplings between distant quantum dots offer a potential solution to
these challenges.  Recently, hybrid superconductor-semiconductor devices have
been put forward, demonstrating coherent coupling between a quantum dot and a
microwave field confined in a superconducting resonator
\cite{bruhat_strong_2016,stockklauser_strong_2017,mi_strong_2017}.
Conceptually, such large-scale resonators can be used to couple spatially
separated qubits.

Such a hybrid solution has to pair different technologies and an all-electronic
solution on chip is highly sought after. Examples of the latter include
RKKY-mediated coupling between two dots connected by a large open dot
\cite{craig_tunable_2004} and proposals for coupling quantum dots via
edge-modes in the quantum Hall regime
\cite{yang_quantum-hall_2002,scarola_pseudospin_2003,elman_2017}. An
alternative novel approach involves the introduction of an electronic cavity
as a mediator of long-distance coupling: recently, such an electronic cavity
that sustains coherent fermionic modes
\cite{katine_point_1997,hersch_diffractive_1999} has been strongly coupled to
a quantum dot \cite{rossler_transport_2015}. The distinct spin-coherent
signatures observed in this dot--cavity setup have spurred further
theoretical~\cite{ferguson2016transport,
ferguson_long-range_2017,ulloa_conductance_2017} and experimental
~\cite{Yan2017, steinacher2017scanning} work and motivates its use as a
quantum bus.

In this work, we demonstrate tunable coherent coupling between distant
quantum dots in a mesoscopic semiconducting architecture. Using a novel kind
of electronic cavity that sustains whispering-gallery modes
\cite{raman_whispering-gallery_1921,zhao_creating_2015}, we achieve
suppression of cross-talk between the dots alongside a selective coupling
mechanism. Specifically, we report on four transport spectroscopy experiments
that systematically demonstrate these features. Our device can serve as a
viable technological solution to the scalability challenge of semiconductor
quantum information processors. At the same time, it offers a novel platform
for the investigation of fascinating many-body problems in solid state physics
such as the two-impurity Kondo system \cite{chang_kondo_2009}.

\begin{figure}[t!]
\includegraphics[width=0.95\columnwidth]{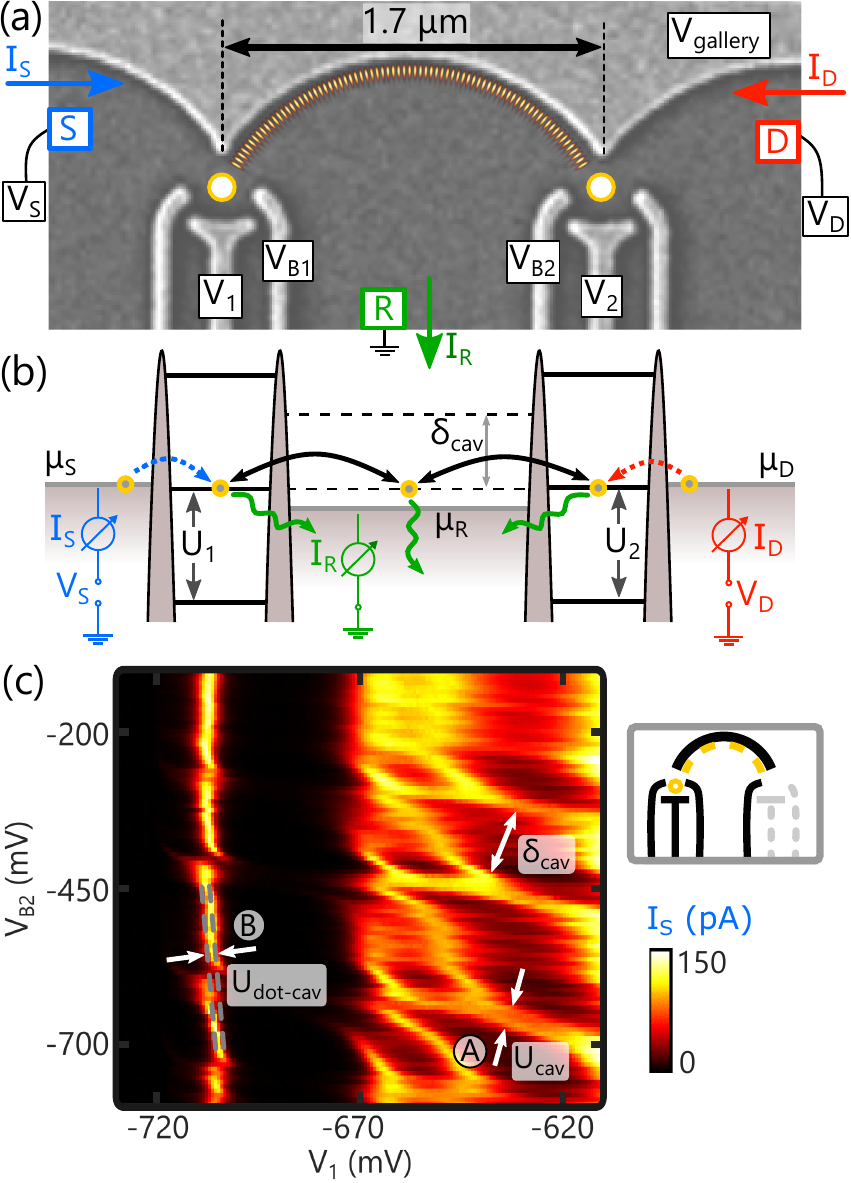}
\caption{\label{fig:sample}(a) Scanning electron micrograph of the device;
bright features are metallic top-gates. Negative voltages can be applied to
all the gates, with labels indicating those explicitly discussed in the text.
The (yellow) circles mark the two quantum dots. A numerically calculated local
density of states map (using KWANT~\cite{groth_kwant_2014}) of a
whispering-gallery cavity mode close to the Fermi energy,
$\varepsilon_F=$~\SI{7.85}{\milli\electronvolt} is shown as an overlay. The
three boxes (S), (D) and (R) label ohmic contacts to the two-dimensional
electron gas.  The contact to lead (R) is grounded at all times, while the
other two are connected to DC voltage sources. The sign of the measured
currents $I_\text{S}$, $I_\text{D}$, and $I_\text{R}$ is positive when
electrons flow in the direction of the arrows. (b) Schematic energy diagram of
the full dot--cavity--dot system including the confined energy levels of the
two dots (black solid lines), the continuous electronic dispersion in the
leads (grey boxes), and the cavity modes (black dashed lines). The colored
arrows indicate possible transport processes in the system. (c) Transport
spectroscopy of the dot--cavity system (experiment I).  Vertical lines
correspond to transport through resonances of the left dot, while the diagonal
lines are identified as cavity modes. Avoided crossings between these two sets
of features are evidence of dot--cavity
hybridization~\cite{rossler_transport_2015, ferguson2016transport,
ferguson_long-range_2017}. The sketch on the right indicates the gates used
for this experiment. The width of the cavity levels, indicated by (A) gives an
upper bound on the cavity charging energy, while the offset indicated by (B)
sets an upper bound on the dot--cavity electrostatic interaction.}
\end{figure}

Our experiments are conducted using different configurations of the device
shown in Fig.~\ref{fig:sample}(a).  The device is cooled to an electronic
temperature of $\sim$\SI{24} {\milli\kelvin} in a dilution refrigerator.  It
is composed of a \TDEG\ that resides \SI{90}{\nano\metre} underneath the
surface of a GaAs/AlGaAs heterostructure, where lithographically defined
metallic top gates act as Schottky contacts.  Applying suitable negative
voltages depletes the underlying \TDEG\ to form two spatially separated quantum
dots set \SI{1.7}{\micro\metre} apart from each other. Each dot is confined
using three finger gates and a large arc-shaped gate (dubbed gallery). In
the experiments below, we apply the same fixed voltage $V_\text{gallery}$ to
the gallery gate, ensuring depletion of the \TDEG\ and contributing to the
electrostatic definition of the dots. The special feature of our device is the
presence of a quasi-1D electronic cavity that sustains coherent
whispering-gallery modes, as verified by KWANT simulations~\cite{groth_kwant_2014}, see Fig.~\ref{fig:sample}(a).  These modes are
spin-degenerate states embedded within the Fermi sea, i.e., electron screening
effectively removes charging effects on the cavity.  Therefore, the cavity
modes modulate the local density of states of the central reservoir and couple the two dots.  Figure~\ref{fig:sample}(b) shows a
schematic energy diagram of the full dot--cavity--dot system with the relevant
transport processes between the setup's constituents.

\textit{In experiment I}, we demonstrate the spin-coherent coupling of the
electronic cavity and the left dot. The energy of the dot is controlled by the
plunger gate voltage $V_\text{1}$, while the cavity is defined on the left by
the dot and on the right by $V_\text{B2}$, which additionally can tune the
length of the cavity. We perform equilibrium transport spectroscopy of this
system \cite{rossler_spectroscopy_2014} as a function of both $V_\text{1}$ and
$V_\text{B2}$, thus tuning the dot and cavity levels while having a small
bias-voltage between reservoirs (S) and (R). The result of this experiment is
presented in Fig.~\ref{fig:sample}(c), where we observe signatures of a
competition between a dot--cavity singlet formation and Kondo transport, 
similar to Refs.\ \cite{rossler_transport_2015,ferguson_long-range_2017}.
This result confirms that we have successfully created a coherent fermionic
cavity in a novel whispering-gallery mode geometry. Similar experiments
confirm the coherent dot--cavity coupling of the second dot.

\begin{figure*}[t]
\includegraphics[width=0.95\textwidth]{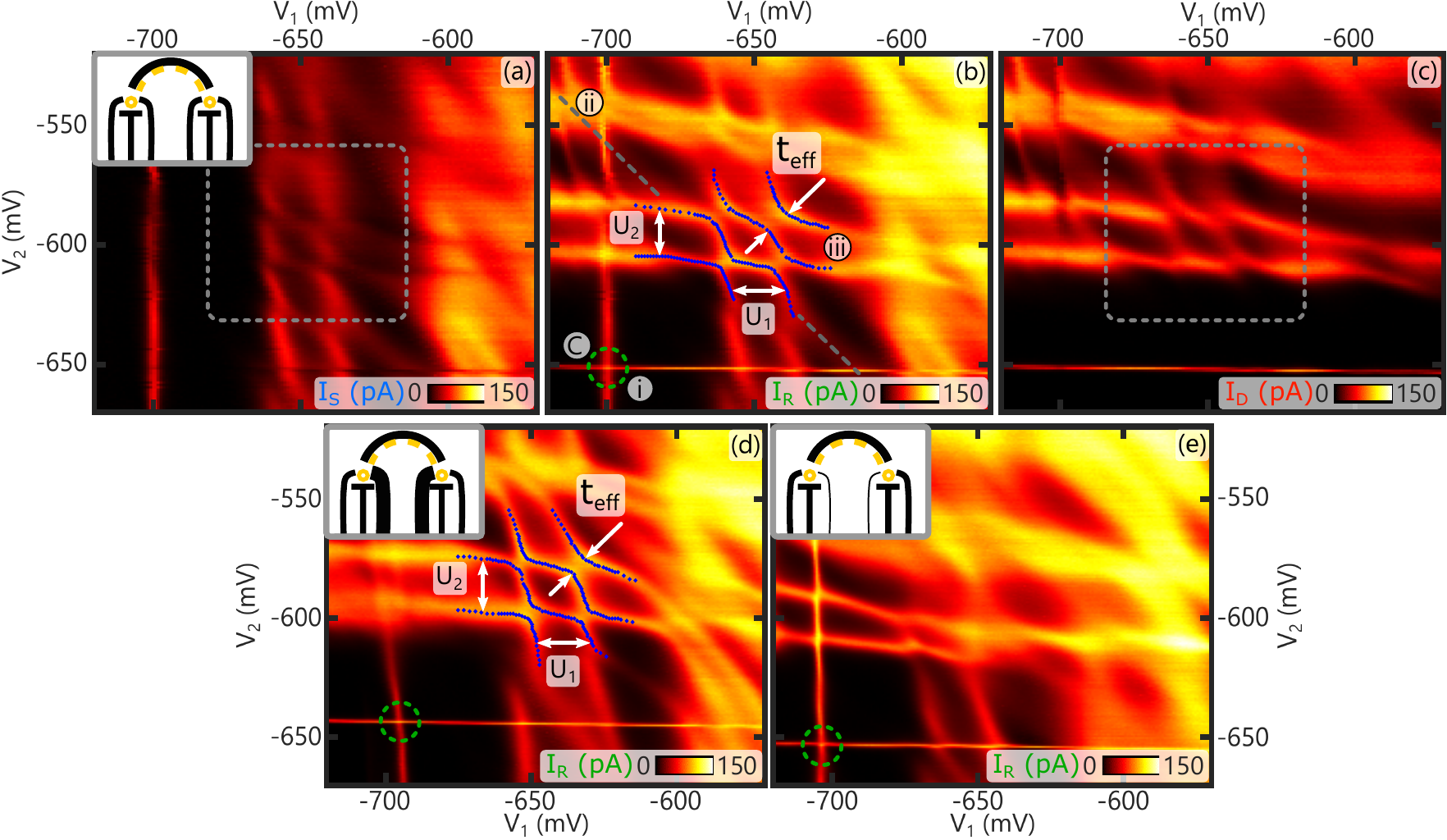}
\caption{\label{fig:Coupled}(a-c) Spectroscopy of the dot--cavity--dot system
exhibiting long-distance coupling between the two dots (experiment II, labels
are color coded as in Fig.~\ref{fig:sample}). The (grey) boxes in (a) and (c)
indicate dot--cavity--dot coupling-related features, corresponding to
transport category (iii) (see discussion in the text), i.e., avoided crossings
between vertical and horizontal resonances. The charging energy of the two
dots $U_i$ ($i = 1,2$) is highlighted in (b). The (blue) points correspond to
the peak position of the Coulomb resonances. Several avoided crossings appear, one of
them indicated by the (white) arrows. The (green) dashed circle in the
bottom-left corner belongs to transport category (i) where the two
resonances associated with the left and right dot are decoupled [with the
individual resonances visible in (a) and (c), respectively]. This crossing constrains
the inter-dot charging energy to be negligibly small (C). The dashed (grey) line
refers to a transport signature of type (ii).  (d,e) Transport measurements of
the same system as in experiment II, but with asymmetric tunnel barriers
(experiment III). Weaker (d) and stronger (e) coupling regimes of the
dot--cavity--dot hybrid are probed [symbols refer to the same quantities as
in (b)]. The upper-left insets sketch which gates are active for each
measurement.}
\end{figure*}

Having established the existence of an electronic cavity that can couple to
our dots, we formulate a simple dot--cavity--dot model for the full device,
see Fig.~\ref{fig:sample}(b) \cite{supmat}. Generally, electrons from each dot
can (i) directly tunnel to (R), (ii) tunnel-couple independently to the
cavity, forming a dot--cavity hybrid state, or (iii) form a dot--cavity--dot hybrid
state with a wavefunction spanning both dots and the cavity, see Fig.\
\ref{fig:Coupled} for examples of such transport signatures. Tracing out the
cavity in case (iii), an effective dot--dot tunnel-coupling is achieved which depends on the energetic configuration of
the dot--cavity--dot system. Aligning the cavity-, dot-, and Fermi-levels, the
effective coupling is equal to the dot--cavity tunnel amplitude
\(t_{\mathrm{eff}}=t_{\mathrm{dot-cav}}\); detuning the cavity level by
$\varepsilon_{\mathrm{cav}}$, a perturbative analysis~\cite{supmat} provides
the reduced effective coupling \(t_{\mathrm{eff}}\sim t_{\mathrm{dot-cav}}^2/
\varepsilon_{\mathrm{cav}} \). Note, that significantly different dot--cavity
couplings for the two dots leads to case (ii) and a suppression of
$t_{\mathrm{eff}}$. We show below that electrostatic interactions between
electrons in different dot/cavity elements are negligible.

\textit{In experiment II}, we demonstrate the long-distance coherent coupling
between the neighboring dots via the cavity. We study transport through the
dot--cavity--dot system as a function of the two plunger gate voltages
$V_\text{1}$ and $V_\text{2}$ and in response to a small bias-voltage applied
simultaneously to both source (S) and drain (D) reservoirs relative to (R). In
Figs.~\ref{fig:Coupled}(a-c), the currents measured in (S), (R), and (D) are
plotted, respectively. In Figs.~\ref{fig:Coupled}(a) and (c), the measurement
of the source ($I_{\rm\scriptscriptstyle S}$) and drain currents
($I_{\rm\scriptscriptstyle D}$) allows us to distinguish between the transport
across the individual dots. On the other hand, the measurement of the
reservoir current $I_{\rm \scriptscriptstyle R} = I_{\rm\scriptscriptstyle S}+
I_{\rm\scriptscriptstyle D}$ in Fig.\ \ref{fig:Coupled}(b) emphasizes the
avoided crossings associated with the coherent interdot transport.

The dot--cavity--dot hybrid is modified when changing the plunger gate
voltages, giving us access to all three transport categories (i), (ii), and
(iii) introduced above.  Transport category (i) is clearly seen in the
bottom-left corner of Fig.~\ref{fig:Coupled}(b), where  vertical (resp.
horizontal) resonance lines meet in a right angle (green dashed circle).
Comparing how this feature appears in Figs.~\ref{fig:Coupled}(a) and (c), we
observe that the vertical (resp.  horizontal) line results from independent
transport through the left (right) dot.
Due to the existence of many cavity levels that couple to each of the dots
independently, we observe superimposed signatures of transport category (ii),
see Figs.~\ref{fig:Coupled}(a-c) and compare to Fig.~\ref{fig:sample}(c).
Diagonal lines in Figs.~\ref{fig:Coupled}(a-c) ~\cite{rossler_transport_2015}
are due to coherent coupling of a single dot to the cavity. The grey dashed
lines in Fig.~\ref{fig:Coupled}(b) can be split into left and right transport
features by comparing it to the Figs.~\ref{fig:Coupled}(a) and (c). 

The appearance of transport signatures that resemble a double-dot
charge stability diagram~\cite{van_der_wiel_electron_2002} in all of the three
measured currents is clear evidence for the strong coupling between the dots, i.e., for transport category (iii).
Most importantly, this coupling is mediated by the electronic cavity as will be further verified in Experiment IV~\cite{supmat}.
From the the magnitude of the avoided-crossing gap in Fig.\
\ref{fig:Coupled}(b), we can derive the large dot--dot effective
tunnel-coupling $t_\text{eff}\sim$~\SI{480}{\micro\electronvolt}.
%
%

There are three types of electrostatic interaction that potentially contribute
to the size of the gaps we have associated to the dot--cavity--dot state, (A)
the intracavity charging energy $U_\text{cav}$, (B) the mutual charging energy
$U_\text{dot-cav}$ between the dot and the cavity, and (C) the mutual dot--dot
charging energy $U_{12}$. For the latter, the clean intersection between dot
resonances, see the green dashed circles in Fig.~\ref{fig:Coupled}(b), limits
$U_{12}$ to a pixel-wide avoided crossing of $\sim$ \SI{10}
{\micro\electronvolt}~\cite{ihn2010semiconductor}.  An upper limit for contributions (A) and (B) is
obtained by further investigation of Fig.\ \ref{fig:sample}(c): the injection
of successive electrons into the cavity only negligibly shifts the Coulomb
peaks of the dot, limiting $U_\text{dot-cav}$ to
$\sim$\SI{20}{\micro\electronvolt}; the charging energy of the cavity should
be visible as two parallel sets of cavity modes shifted with respect to each
other by the Coulomb interaction. We find an upper bound of $U_\text{cav}
\simeq$~\SI{30}{\micro\electronvolt}, given by the finite linewidth of the
cavity resonances. All these contributions combined amount to at most
$\sim$\SI{10}{\percent} of the measured coupling energy and we conclude that
the gap opening is dominated by coherent tunneling. In summary, we have shown
that a coherent dot--dot coupling can be mediated between distant dots using
an electronic cavity~\cite{comm_spin}.

It is helpful to place the measured tunnel-coupling $t_\text{eff}
\sim$~\SI{480} {\micro\electronvolt} into context with (other) typical device
parameters, e.g., the charging energy of the two dots $U_\text{1} \simeq
U_\text{2} \simeq$~\SI{1}{\milli\electronvolt} (highlighted in
Fig.~\ref{fig:Coupled}(b)), the dot single-particle level spacing
$\sim$\SI{400}{\micro eV}, as well as coupling energies obtained in standard
double-dot experiments. For the latter, typical tunnel splittings amount to
$\sim$\SI{100}{\micro\electronvolt} \cite{blick_formation_1998}, with the
total coupling energy including comparable tunnel and electrostatic
contributions \cite{hatano_electron-spin_2004,hatano_single-electron_2005}.
The separation between the dots in the experiments of Refs.\
\cite{hatano_electron-spin_2004,hatano_single-electron_2005} is of the
order of a few hundred nanometers, while in our sample, the distance between
the dots is almost \SI{2}{\micro\metre}. Hence, in spite of the larger
separation between dots, the cavity-assisted coupling mechanism studied in the
present work provides comparable effective coupling strengths between dots,
with the additional advantage of a greatly suppressed electrostatic
cross-talk.

\textit{In experiment III}, we report on the gate-tunability of our setup. We
establish weak and strong coupling regimes by changing the tunnel barriers
(and hence the coupling) of the dots towards the cavity via the voltages
$V_\text{B1}$ and $V_\text{B2}$ [Fig.~\ref{fig:sample}(a)]. In the weak coupling regime
[Fig.~\ref{fig:Coupled}(d)], the barriers confine the dots more strongly,
resulting in narrower avoided crossings and a reduced coupling $t_\text{eff}
\simeq$~\SI{330}{\micro eV}, a $\sim$\SI{31}{\percent} reduction of with
respect to the situation in Fig.~\ref{fig:Coupled}(b).  In the strong coupling
configuration [Fig.~\ref{fig:Coupled}(e)], the reduced confinement of the dots
washes out the signatures of strong dot--cavity--dot hybridization that was
observed in Fig.~\ref{fig:Coupled}(c). The coupling between the dots and the
cavity increases to the point where we expect the system to behave more like a
single quantum dot with a large area, instead of three separated (but coupled)
systems. Experiment III offers a path towards achieving complete On/Off
switching of the coupling with future improved designs.

\begin{figure}[t]
\includegraphics[width=0.95\columnwidth]{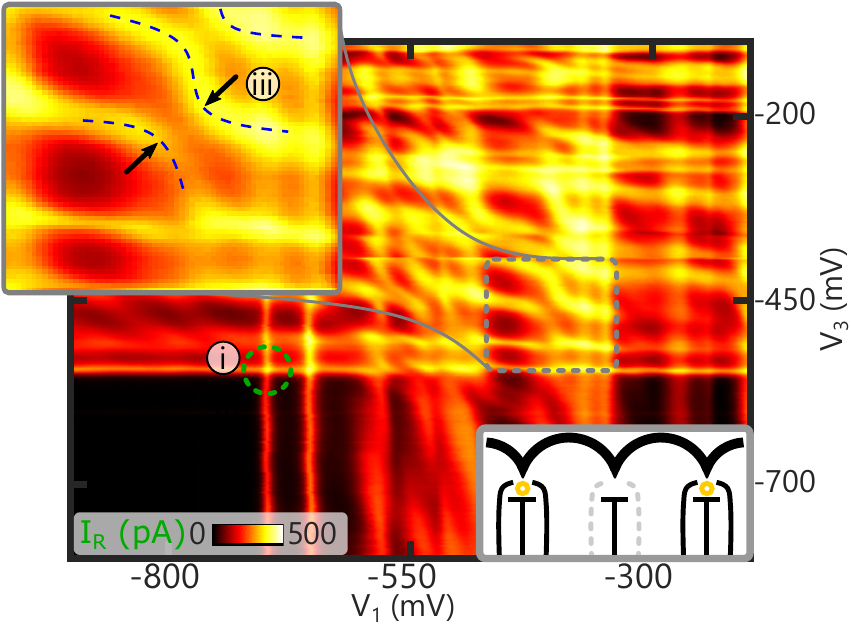}
\caption{\label{fig:NN_QD} Transport through a hybrid system consisting of a longer-ranged electronic cavity that couples next-nearest-neighbor quantum dots (experiment IV). The bottom-right inset schematically
shows the gates configuration for the experiment~\cite{supmat}. Similarly to
experiment II, we find horizontal and vertical type (i) resonances due to
separate transport through the energy levels of the two dots. Avoided
crossings of type (iii), appear at the intersection between specific
resonances, see the magnified upper-left inset (black arrows). These are clear signature for coherent
dot--cavity--dot coupling~\cite{supmat}. The (blue) dashed lines are guides to the eye.}
\end{figure}

\textit{In experiment IV}, we test the potential of the cavity-mediated
coupling to extend over longer distances. Taking advantage of a third dot
present in our sample~\cite{supmat}, we repeat experiment II with a
next-nearest-neighbor configuration (see lower inset in Fig.~\ref{fig:NN_QD}).
The dots in this case reside \SI{3.5}{\micro\metre} apart from each other and
are coupled via a whispering-gallery mode spanning two cavity arcs; the gate
controlled by $V_\text{2}$ partially depletes the central reservoir without
forming a dot and connects the two cavities~\cite{supmat}. In
Fig.~\ref{fig:NN_QD}, we observe avoided crossings characteristic of transport
category (iii), as highlighted in the upper inset. Along with measurements of
the individual source and drain currents \cite{supmat}, we obtain clear
evidence of a dot--dot cavity-mediated tunnel coupling. The splitting is
observed only when the two-arc gallery mode is formed, i.e., it can be
switched using the gate bias $V_\text{2}$~\cite{supmat}.

The long-distance, cavity-mediated tunnel-coupling between dots investigated
in our device offers an all-electronic controllable platform for quantum
information processing. In particular, the reduced electrostatic cross-talk
and the tunable long-distance coherent coupling provide a possible solution to
the scalability challenges in semiconducting architectures. Further
improvements of our prototypical cavity-design may offer higher-control and
selective connectivity, and thus the possibility of entanglement experiments
for two or more qubits \cite{mehl_two-qubit_2014,srinivasa_tunable_2015}.
Furthermore, our system can be used to study many-body physics phenomena, such
as exotic Kondo systems involving two magnetic impurities, i.e., two isolated
spins confined in the dots \cite{chang_kondo_2009}, the competition between
cavity-assisted and RKKY-mediated coupling \cite{craig_tunable_2004}, as well
as, the realization of a potential Kondo-cat
state~\cite{ferguson_long-range_2017}.

\begin{acknowledgments}
We thank T.\ Kr\"ahenmann and G.\ Burkard for illuminating discussions. We acknowledge support from the ETH FIRST laboratory and financial support from the Swiss National Science Foundation, Division 2 and through the National Centre of Competence in Research "QSIT - Quantum Science and Technology".
\end{acknowledgments}

\title{Cavity-mediated coherent coupling between distant quantum dots}

\author{Giorgio\ Nicol\'i}
	\affiliation{Solid State Physics Laboratory, ETH Z\"urich, 8093 Z\"urich, Switzerland}

\author{Michael Sven\ Ferguson}
	\affiliation{Institute for Theoretical Physics, ETH Z\"urich, 8093 Z\"urich, Switzerland}

\author{Clemens\ R\"ossler}
	\affiliation{Infineon Technologies Austria, Siemensstra{\ss}e 2, 9500 Villach, Austria}
			
\author{Alexander\ Wolfertz}
	\affiliation{Solid State Physics Laboratory, ETH Z\"urich, 8093 Z\"urich, Switzerland}

\author{Gianni\ Blatter}
	\affiliation{Institute for Theoretical Physics, ETH Z\"urich, 8093 Z\"urich, Switzerland}
	
\author{Thomas\ Ihn}
	\affiliation{Solid State Physics Laboratory, ETH Z\"urich, 8093 Z\"urich, Switzerland}
	
\author{Klaus\ Ensslin}
	\affiliation{Solid State Physics Laboratory, ETH Z\"urich, 8093 Z\"urich, Switzerland}
	
\author{Christian\ Reichl}
	\affiliation{Solid State Physics Laboratory, ETH Z\"urich, 8093 Z\"urich, Switzerland}
	
\author{Werner\ Wegscheider}
	\affiliation{Solid State Physics Laboratory, ETH Z\"urich, 8093 Z\"urich, Switzerland}

\author{Oded\ Zilberberg}
	\affiliation{Institute for Theoretical Physics, ETH Z\"urich, 8093 Z\"urich, Switzerland}

\newpage
\cleardoublepage
\setcounter{figure}{0}
\renewcommand{\figurename}{Supplementary Information Figure}

\onecolumngrid
\begin{center}
\textbf{\normalsize Supplemental Material for:}\\
\vspace{3mm}
\textbf{\large Cavity-mediated coherent coupling between distant quantum dots}
\vspace{4mm}

{ Giorgio\ Nicol\'i,$^{1}$ Michael Sven\ Ferguson,$^{2}$, Clemens\ R\"ossler,$^{1,3}$ Alexander\ Wolfertz,$^{1}$ Gianni\ Blatter,$^{2}$\\ Thomas\ Ihn,$^{1}$ Klaus\ Ensslin,$^{1}$ Christian\ Reichl,$^{1}$ Werner\ Wegscheider,$^{1}$ and Oded Zilberberg,$^{2}$}\\
\vspace{1mm}
\textit{\small $^{1}$Solid State Physics Laboratory, ETH Z\"urich, 8093 Z\"urich, Switzerland\\
$^{2}$Institute for Theoretical Physics, ETH Z\"urich, 8093 Z\"urich, Switzerland\\
$^{3}$Infineon Technologies Austria, Siemensstra{\ss}e 2, 9500 Villach, Austria}

\vspace{5mm}
\end{center}
\setcounter{equation}{0}
\setcounter{figure}{0}
\setcounter{table}{0}
\setcounter{page}{1}
\makeatletter
\renewcommand{\theequation}{S\arabic{equation}}
\renewcommand{\bibnumfmt}[1]{[S#1]}
\renewcommand{\citenumfont}[1]{S#1}

\twocolumngrid

\section{Sample fabrication and description}

Our device was fabricated from a GaAs/AlGaAs modulation-doped, single-interface heterostructure grown by molecular-beam epitaxy (MBE). In this way a two-dimensional electron gas (2DEG) is confined in a triangular potential well in the vertical (growth) direction. The 2DEG is localized in a GaAs layer close to the interface with a GaAlAs layer, and \SI{90}{nm} below the surface. The sheet electron density of the wafer is $n_\mathrm{s} =$~\SI{2.2e11}{cm^{-2}} and the mobility is $\mu_\mathrm{e} =$~\SI{3.4e6}{cm^2/(Vs)}, both measured at \SI{1.3}{\kelvin} with standard magnetotransport techniques.

The base heterostructure was additionally processed to define the desired nanostructures. Using optical lithography, we define ohmic contacts to the electron gas and gate leads. In a final step, we define the finer gate structures using electron-beam lithography. The lithography steps are accompanied by etching or metal evaporation and lift-off processes, and in the case of the ohmic contacts thermal annealing.

\begin{figure*}[h]
	\includegraphics[width=0.95\textwidth]{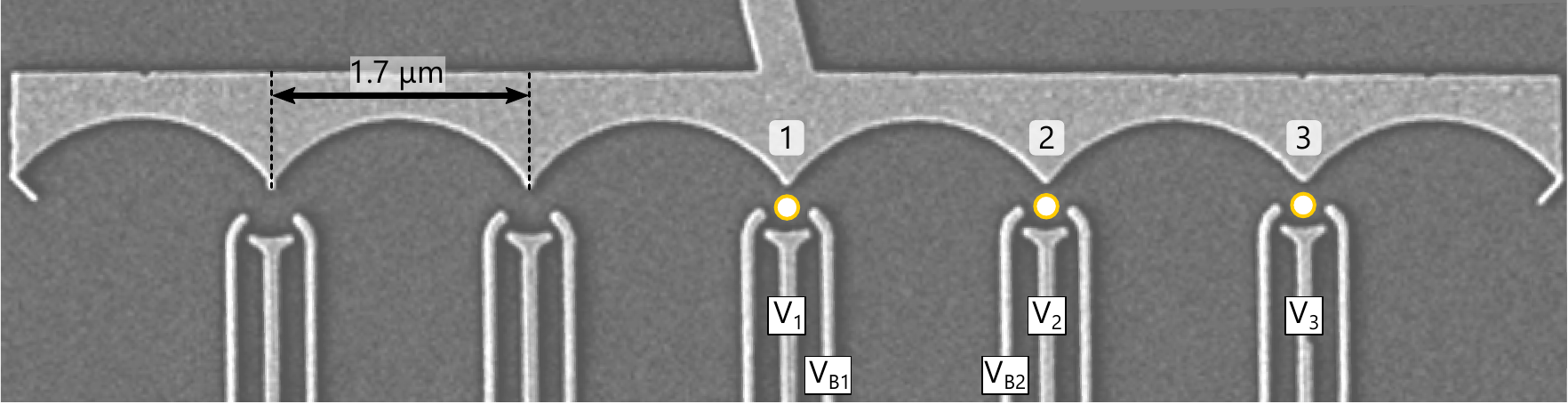}
	\caption{\label{fig:sample_supmat} Scanning electron micrograph of our device. The picture shows the part of the sample with metallic top gates defined by electron-beam lithography. Gate leads and ohmic contacts defined with optical lithography are outside the scanned area.}
\end{figure*}

An SEM picture of the device is shown in Fig.~\ref{fig:sample_supmat}. Five spatially-separated quantum dots can be defined with the metallic top-gates. The dots labeled 1 and 2 are those used for experiments I - III (see main text). The additional dot mentioned in the discussion of experiment IV (main text), is the one labeled 3 in Fig.~\ref{fig:sample_supmat}. In the main text, we report on experiments conducted on the labeled dots (1,2 and 3). Similar results were obtained using other dot configurations.

\section{Robustness of the long-distance cavity-mediated coherent coupling}
\begin{figure*}[h]
	\includegraphics[width=0.95\columnwidth]{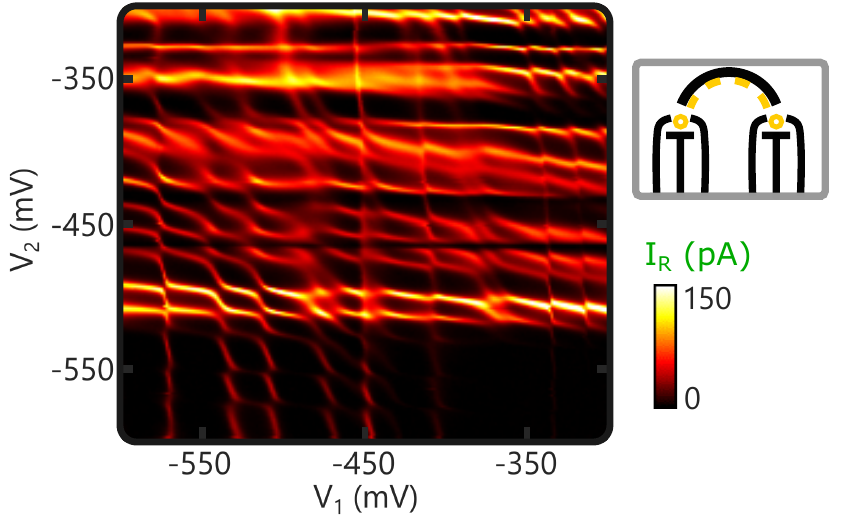}
	\caption{\label{fig:coupled_supmat} Equilibrium transport spectroscopy of neighboring coupled quantum dots (dots 1 and 2). Vertical and horizontal lines correspond to transport through the Coulomb resonances of left and right dot, respectively. Multiple avoided crossings between the two sets of resonances are observed, which are signatures of a dot-cavity-dot hybrid. The cavity levels are kept at constant energies by compensating the \(V_1,V_2\) induced shift by an equal and inverse $V_\mathrm{B1},V_\mathrm{B2}$ shift such that the avoided crossings are observed over a large fraction of the phase-space. The sketch on the right indicates the gates biased for this experiment.}
\end{figure*}

In the main text, the cavity-mediated effective dot--dot coupling extends over a small fraction of the \(V_1,V_2\) phase space. This is a result of the cavity levels being shifted by the change of these voltages. We can significantly increase the phase space fraction by performing linear voltage compensation. The gates which tune the tunnel couplings $V_\mathrm{B1},V_\mathrm{B2}$ are used to induce an opposite cavity shift to the one caused by the voltages \(V_1,V_2\). This keeps the cavity levels, \(\epsilon_{{\rm c}\sigma}\) in the model below~\ref{sec:theory}, constant throughout a much larger fraction of the \(V_1,V_2\) phase space where we thus observe the dot--dot avoided crossings. This can be seen in Fig.~\ref{fig:coupled_supmat}, where we performed this experiment in dots 1 and 2.

\section{Supplementary data for experiment IV}
\begin{figure*}[h]
	\includegraphics[width=0.95\textwidth]{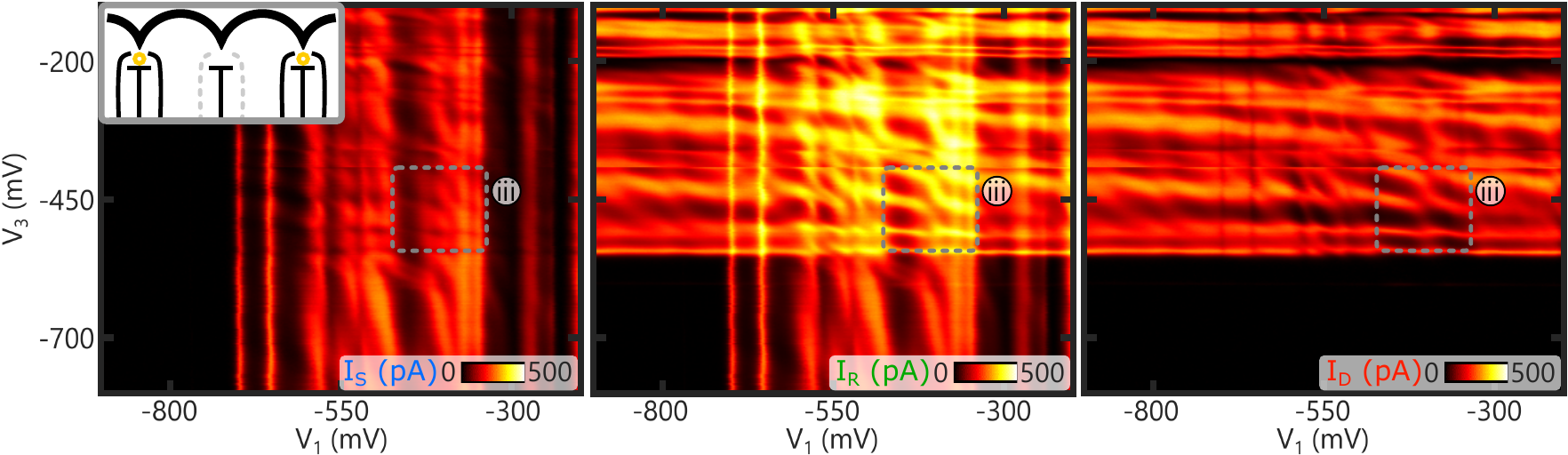}
	\caption{\label{fig:next_neigh_qd_supmat} Spectroscopy of the next-nearest-neighbor quantum dots system (experiment IV, dot 1 and 3).
	In the grey dashed box, we identify signatures of coherent transport across the entire device [transport category (iii)]. Namely the avoided crossings in \(I_\mathrm{R}\) are accompanied by vertical (horizontal) transport features in the source $I_\mathrm{S}$ (drain $I_\mathrm{D}$) current and thus corresponds to the formation of a dot--cavity--dot hybrid.}
\end{figure*}

In experiment IV, we analyze the coupling of next-nearest-neighbor quantum dots (labeled 1 and 3). In this case the cavity spans two arcs of the gallery gate connected by a channel defined by the gate \(V_2\). Notably, we still observe signatures of coherent dot--cavity--dot hybrid states testifying to the robustness of our setup. In order to properly identify these signatures we consider the source $I_\mathrm{S}$, drain $I_\mathrm{D}$ and reservoir currents $I_\mathrm{R}=I_\mathrm{S} + I_\mathrm{D}$ (see Fig.~\ref{fig:next_neigh_qd_supmat}). The avoided crossings in \(I_\mathrm{R}\) are accompanied by vertical (horizontal) transport features in the source $I_\mathrm{S}$ (drain $I_\mathrm{D}$) current. These clearly indicate that changes in one dot are influencing the other and that we have thus set up a coherent hybrid dot--cavity--dot molecule over the double the spatial extent of experiment III.

\begin{figure*}[h]
	\includegraphics[width=0.95\textwidth]{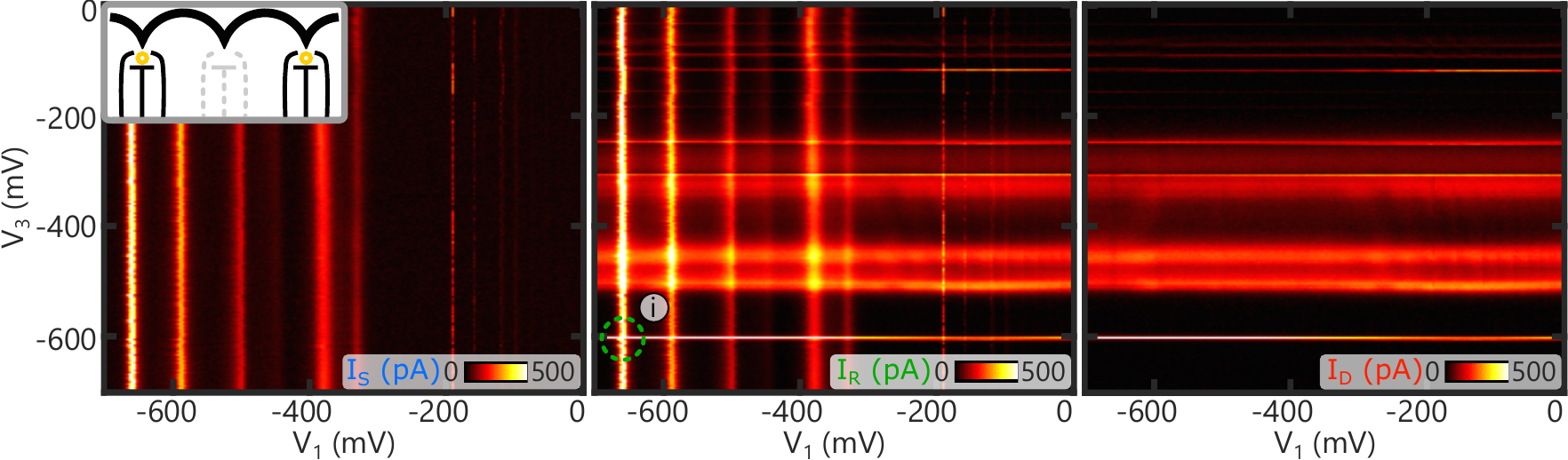}
	\caption{\label{fig:next_neigh_qd_supmat_nocav} Spectroscopy of the next-nearest-neighbor quantum dots system (experiment IV, dot 1 and 3) with gate $V_\mathrm{2}$ grounded. In this case, only transport category (i) is observed. The green dashed circle highlights one of the many clean intersection between the sets of vertical and horizontal resonant lines. These features correspond to electron tunneling through unperturbed dots' resonances, identifiable by comparing the measurement of source ($I_\mathrm{S}$) and drain ($I_\mathrm{D}$) currents (left and right panels, respectively).}
\end{figure*}

We found that slightly tuning the channel gate \(V_2 = -500 \textrm{mV}\) away from its optimal value is sufficient to switch off the effective dot--dot coupling. The extended cavity is then pinched off into two separate cavities. Additionally, we can turn off the effective coupling by grounding $V_\mathrm{2}$ such that the cavity is no longer well defined. The latter case can be seen in Fig.~\ref{fig:next_neigh_qd_supmat_nocav}, where we observe only the superposition between two sets of horizontal (vertical) resonant lines corresponding to an extended region of transport category (i). 

\section{Theoretical Model} \label{sec:theory}
\begin{figure*}[h]
	\includegraphics[width=0.45\textwidth]{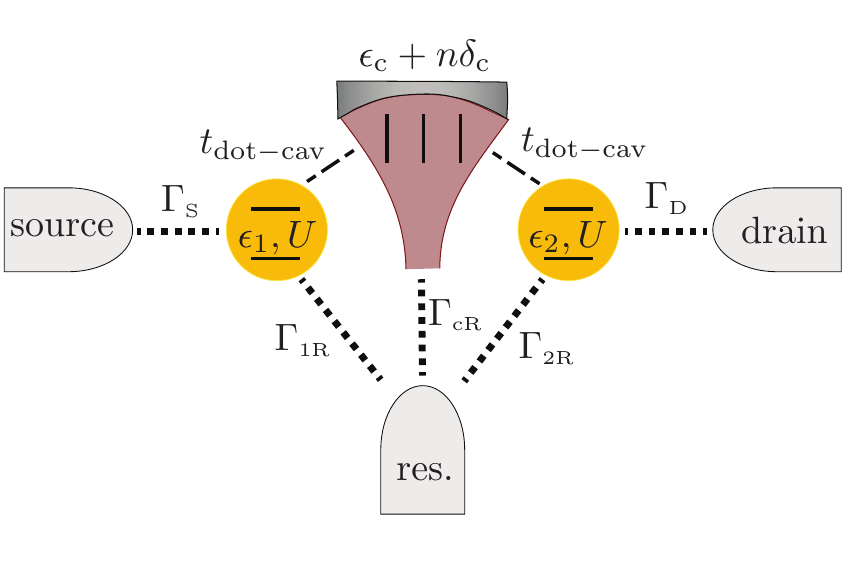}
	\caption{
		\label{fig:supMatModel} 
		A pictorial representation of the effective model. It includes three leads (source, drain, and reservoir), two dots with independent levels \(\epsilon_{1},\epsilon_{2}\), equal onsite interaction \(U\), and a cavity with levels spaced at \(\epsilon_{\rm c}+n\delta_{\rm c}\), where \(n\) is an integer. The rates \(\Gamma\) (dotted lines) weakly couple the dot--cavity--dot molecule to the leads and provide a chemical potential \(\mu\), while \(t_{\rm dot--cav}\) coherently couples the dots to the cavity. Note that the three couplings to the reservoir must be accounted for in such a way as to avoid an artificial Fano interference term~\cite{ferguson_long-range_2017}.
		}
\end{figure*}

We follow the methods laid out in Ref.~\cite{ferguson_long-range_2017} to obtain an effective model describing the dot--cavity--dot device. We first simulate the single particle properties of the cavity using KWANT~\cite{groth_kwant_2014}, we then argue for the removal of Fano interference terms based on the two dimensionality of the system and thus arrive at a model which can be represented pictorially, as in Fig.\ref{fig:supMatModel}.
 
This model Hamiltonian reads
\begin{align}
H = H_{\rm molecule} + H_{\rm leads} + H_{\rm coupling}\,,
\end{align}
where the molecular Hamiltonian,
\begin{align}
H_{\rm molecule} = H_{\rm dots} + H_{\rm cav} + H_{\rm tun}\,,
\end{align}
describes the dot--cavity--dot system. It is in turn split into a dot contribution
\begin{align}
H_{\rm dots} = \sum_{\alpha,i,\sigma} 
\epsilon_{\alpha i \sigma} d_{\alpha i \sigma}^\dagger d_{\alpha i \sigma}^\pdagger
+ \sum_{\alpha}U_\alpha \frac{n_{\alpha}^2-n_{\alpha}}{2}\,,
\end{align}
where \(\alpha={1,2}\) indexes the two different dots, \(i\in\mathbb{N}\) indexes the states in the given dot, and \(\sigma = \uparrow,\downarrow\) indexes the spin. The energy \(\epsilon_{\alpha i \sigma}\) gives the energy of the corresponding, \(\alpha,i,\sigma\) state while \(U\) is the interaction energy of the dots (taken to be equal in each dot). Additionally we have introduced the fermionic creation and annihilation operators \(d_{\alpha i \sigma}^\dagger,d_{\alpha i \sigma}^\pdagger\) corresponding to the state \(\alpha,i,\sigma\), and the total number of particles in a given dot is 
\begin{align}
n_\alpha = \sum_{i,\sigma}d_{\alpha i \sigma}^\dagger d_{\alpha i \sigma}^\pdagger\,.
\end{align}
The cavity is assumed to be a set of energetically equidistant non-interacting fermionic levels and are thus described by
\begin{align}
H_{\rm cav} = 
\sum_{n,\sigma} (\epsilon_{{\rm c}\sigma} + n\delta_{\rm c})c_{n\sigma}^\dagger c_{n\sigma}^\pdagger\,,
\end{align}
where \(n=0,\pm 1, \pm 2,...\), up to a cutoff where the equidistant approximation is no longer valid. We have also introduced the cavity level spacing \(\delta_{\rm c}\), the energy of the zeroth cavity level \(\epsilon_{{\rm c},\sigma}\) and the cavity creation and annihilation operators \(c_{n\sigma}^\dagger,c_{n\sigma}^\pdagger\).
The final piece of the artificial molecule Hamiltonian is the tunneling term between the two dots and the cavity
\begin{align}
H_{\rm tun} =  \sum_{\alpha,i,n,\sigma}
\Omega_\alpha d_{\alpha,i,\sigma}^\dagger c_{n,\sigma}^\pdagger
+h.c.\,,
\end{align}
where \(h.c.\) stands for Hermitian conjugate and the tunneling amplitude \(\Omega_\alpha\) is a function of which dot \(\alpha\) the tunneling involves only. 

The \(H_{\rm leads}\) and \(H_{\rm coupling}\) terms describe three leads coupled directly to either the dots or the cavity with energy independent tunneling amplitudes. For the purpose of calculating the ground state of the molecular Hamiltonian they are unimportant as they serve only to provide a chemical potential \(\mu\equiv0\) which we define as the zero point of energy.

As there is no magnetic field in the experiment we immediately have \(\epsilon_{\alpha i\sigma} \equiv \epsilon_{\alpha i}\) and \(\epsilon_{{\rm c}\sigma}\equiv\epsilon_{\rm c}\). For simplicity we restrict ourselves to the case of a single state in each dot \(\epsilon_{\alpha,i,\sigma}\equiv\epsilon_{\alpha}\) and the cavity \(\delta_{\rm c}\to\infty\) such that there are only 64 states in the Hilbert space which can be easily diagonalized numerically to find good agreement with the experiment. The bottom-left to top-right diagonal in Fig. 2 of the main text, along which the avoided crossings associated with coherent coupling of the two dots occur corresponds to the case \(\epsilon_{1}=\epsilon_2\). This case can be treated analytically to extract the gap sizes. Writing down the Hamiltonians in their matrix representation following~\cite{ferguson_long-range_2017}, we immediately find that there are two principal regimes: when the cavity level is resonant with the Fermi energy \(\epsilon_{\rm c} = 0\) we obtain a gap \(\sim\Omega\) (as in degenerate perturbation theory); on the other hand when the cavity is far detuned we obtain a gap \(\sim\Omega^2/\epsilon_{\rm c}\) (as in non-degenerate perturbation theory). A future experiment with a direct handle on the cavity level will be able to explore these scalings.


\begin{thebibliography}{41}
\bibitem [{com()}]{comm_spin}%
  \BibitemOpen
  \href@noop {} {}\bibinfo {note} {Even though the cavity-mediated strong
  coupling between dots can be explained both by spinful and spinless models,
  several indications support the spinful case: the dot-cavity coupling (exp.
  I) exhibits spin-coherent coupling between a dot and a set of cavity levels
  with spacing {$\sim$\SI{190}{\micro\electronvolt}}; in the double dot case
  (exp. II) the cavity signatures appear with similar spacing
  (Fig.{~\ref{fig:Coupled}}(a-c)), which suggest that the same set of cavity
  states is responsible for the coupling observed in experiments I and
  II.}\BibitemShut {Stop}%
\end{thebibliography}
\end{document}